\documentclass[conference]{IEEEtran}
\IEEEoverridecommandlockouts
\usepackage{cite}
\usepackage{amsmath,amssymb,amsfonts}
\usepackage{algorithmic}
\usepackage{graphicx}
\usepackage{textcomp}
\usepackage{xcolor}
\usepackage{booktabs}
\usepackage{multirow}
\usepackage{amsmath} 
\def\BibTeX{{\rm B\kern-.05em{\sc i\kern-.025em b}\kern-.08em
    T\kern-.1667em\lower.7ex\hbox{E}\kern-.125emX}}

\begin{document}

\makeatletter
\newcommand{\conflinea}{IEEE International Conference on Data Mining (ICDM), Workshop on User Modeling and Recommendation (UMRec), Washington, DC, USA, 2025}
\newcommand{\conflineb}{}

\def\ps@IEEEtitlepagestyle{%
  \def\@oddhead{%
    \parbox[t]{\textwidth}{\centering
      \footnotesize \textit{\conflinea}\\[0.50ex]
      \footnotesize \conflineb
    }%
  }%
  \def\@evenhead{\@oddhead}%
  \def\@oddfoot{}%
  \def\@evenfoot{}%
}
\makeatother

\title{Effectiveness of LLMs in Temporal User Profiling for Recommendation}

\author{
\IEEEauthorblockN{Milad Sabouri\IEEEauthorrefmark{1}, Masoud Mansoury\IEEEauthorrefmark{2}, Kun Lin\IEEEauthorrefmark{1}, Bamshad Mobasher\IEEEauthorrefmark{1}}
\IEEEauthorblockA{\IEEEauthorrefmark{1}DePaul University, USA \\
Email: msabouri@depaul.edu, klin13@depaul.edu, mobasher@cs.depaul.edu}
\IEEEauthorblockA{\IEEEauthorrefmark{2}Delft University of Technology, Netherlands \\
Email: m.mansoury@tudelft.nl}
}

\maketitle
\IEEEpubidadjcol

\begin{abstract}
Effectively modeling the dynamic nature of user preferences is crucial for enhancing recommendation accuracy and fostering transparency in recommender systems. Traditional user profiling often overlooks the distinction between transitory short-term interests and stable long-term preferences. This paper examines the capability of leveraging Large Language Models (LLMs) to capture these temporal dynamics, generating richer user representations through distinct short-term and long-term textual summaries of interaction histories. Our observations suggest that while LLMs tend to improve recommendation quality in domains with more active user engagement, their benefits appear less pronounced in sparser environments. This disparity likely stems from the varying distinguishability of short-term and long-term preferences across domains; the approach shows greater utility where these temporal interests are more clearly separable (e.g., Movies\&TV) compared to domains with more stable user profiles (e.g., Video Games). This highlights a critical trade-off between enhanced performance and computational costs, suggesting context-dependent LLM application. Beyond predictive capability, this LLM-driven approach inherently provides an intrinsic potential for interpretability through its natural language profiles and attention weights. This work contributes insights into the practical capability and inherent interpretability of LLM-driven temporal user profiling, outlining new research directions for developing adaptive and transparent recommender systems.
\end{abstract}


\section{Introduction}
Effective recommender systems demand both accurate and transparent suggestions, a challenge exacerbated by the dynamic nature of user preferences. Traditional user profiling, such as averaging item embeddings, often oversimplifies user interests by conflating fleeting short-term desires with stable long-term tastes, hindering both recommendation quality and interpretability. This paper examines the capability of leveraging Large Language Models (LLMs) to capture these temporal dynamics. Our approach generates distinct short-term and long-term textual summaries of user interaction histories, which are then encoded via BERT \cite{devlin-etal-2019-bert} and adaptively fused using an attention mechanism \cite{waswani2017attention} to form a unified user representation.

Our observations reveal a nuanced capability of this LLM-driven approach across varying user engagement patterns. While it tends to improve recommendation quality in active domains like Movies\&\allowbreak TV, with improvements such as 17\% in Recall@10 and 14\% in NDCG@10, benefits are less pronounced in sparser environments like Video Games. Our analysis of these results leads us to hypothesize that this nuanced capability is particularly evident where short-term and long-term preferences are more clearly separable (e.g., Movies\&TV), versus domains with more stable user profiles (e.g., Video Games). This highlights a critical trade-off between LLM-enhanced predictive power and associated computational costs, suggesting context-dependent application.

Also, beyond predictive accuracy, this LLM-driven approach inherently offers an intrinsic potential for transparency. The natural language profiles and learned attention weights provide a pathway toward interpretability, conveying whether recommendations stem from recent or enduring interests. While fully realizing user-facing explanations is a future direction, this work contributes insights into the practical capability and inherent interpretability of LLM-driven temporal user profiling, outlining new research directions for adaptive and transparent recommender systems.

\begin{figure*}[t]
    \centering
    \includegraphics[width=\textwidth]{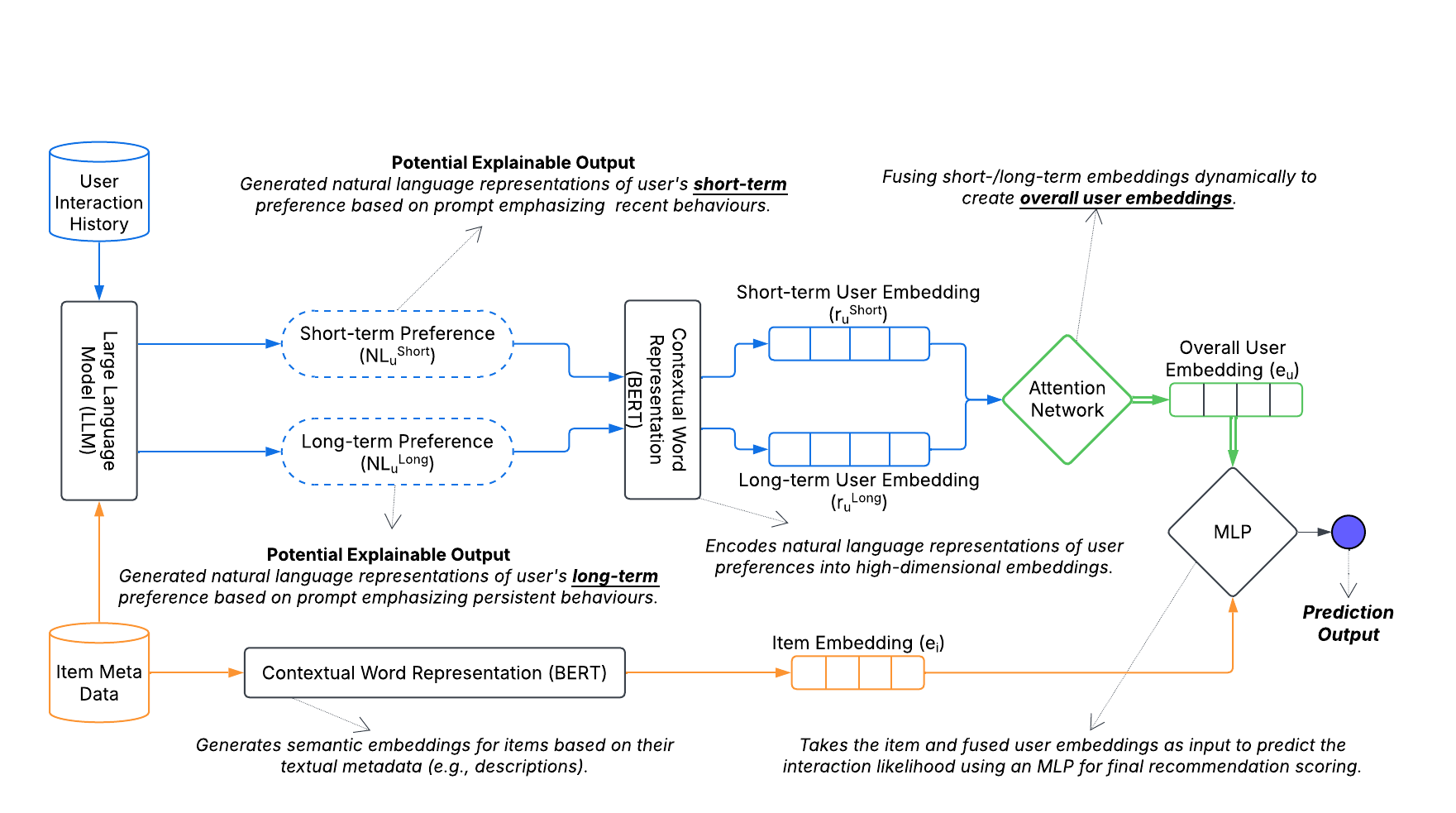}
    \caption{\small Pipeline for LLM-Driven Temporal User Profile Generation and Recommendation.}
    \label{fig:workflow}    
    \vspace{-10pt}
\end{figure*}

\section{Related Work}
Modeling the dynamic nature of user preferences is a long-standing challenge in recommender systems. Traditional methods, such as averaging item embeddings \cite{polignano2021together}, often fail to capture the evolving distinction between short-term and long-term interests. More advanced approaches for temporal and sequential recommendation include time-aware matrix factorization (e.g., TimeSVD++ \cite{koren2009collaborative}), and sequential deep learning models like GRU4Rec \cite{hidasi2015session} and SASRec \cite{kang2018self}, which learn user dynamics from interaction sequences. Recent work has also explored long short-term preference modeling for continuous-time sequential recommendation \cite{chi2022long}.

Research in recommender systems increasingly emphasizes balancing predictive accuracy with interpretability. Early efforts integrated explanations through content summarization or leveraged user-generated reviews \cite{zhang2020explainable}. More recently, Large Language Models (LLMs) \cite{zhao2023survey, chang2024survey} have emerged as powerful tools for generating richer explanations, demonstrating their capacity to enhance interpretability through personalized narratives and collaborative adaptors \cite{zhao2024recommender, wu2024survey, lubos2024llm, ma2024xrec}. Knowledge graph (KG)-based approaches \cite{guo2020survey} have also integrated structured side information for explanations, sometimes combined with LLM outputs \cite{guo2020survey, wang2025knowledge, shimizu2022explainable, shi2024llm}.

Our work intersects these two critical areas by investigating the utility of leveraging LLMs to explicitly model temporal user dynamics for enhanced content-based recommendations and inherent transparency. Unlike existing methods that primarily use static profiles or sequential models without natural language grounding \cite{tan2016improved, kang2018self, zhu2017next}, this study explores an approach that generates distinct natural language summaries for short-term and long-term user behaviors. By encoding and adaptively fusing these semantic summaries, this approach aims to create intrinsically interpretable user profiles. This unique combination transparently highlights the temporal rationale behind recommendations, bridging advanced LLM-based explainability with nuanced temporal dynamics of user interests. This work provides preliminary insights into the practical utility of this novel integration.
\begin{figure*}[t]
    \centering
    \includegraphics[width=\textwidth]{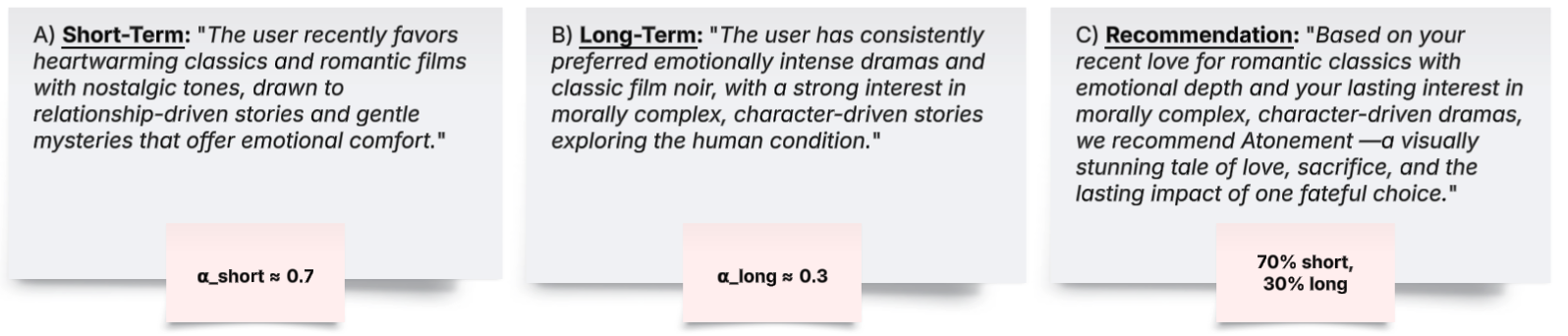}
    \caption{\small A conceptual illustration of the framework’s potential for enhancing transparency. (A) and (B) show the short-term and long-term textual profiles generated and encoded by our current approach. (C) shows a hypothetical extension wherein the final recommendation is explicitly justified by both sets of preferences, further strengthening user-facing transparency—an aspect we plan to explore in future work.}
    \label{fig:case_study}
    \vspace{-10pt}
\end{figure*}

\section{Methodology}

\begin{figure}
    \centering
    \includegraphics[width=0.35\textheight]{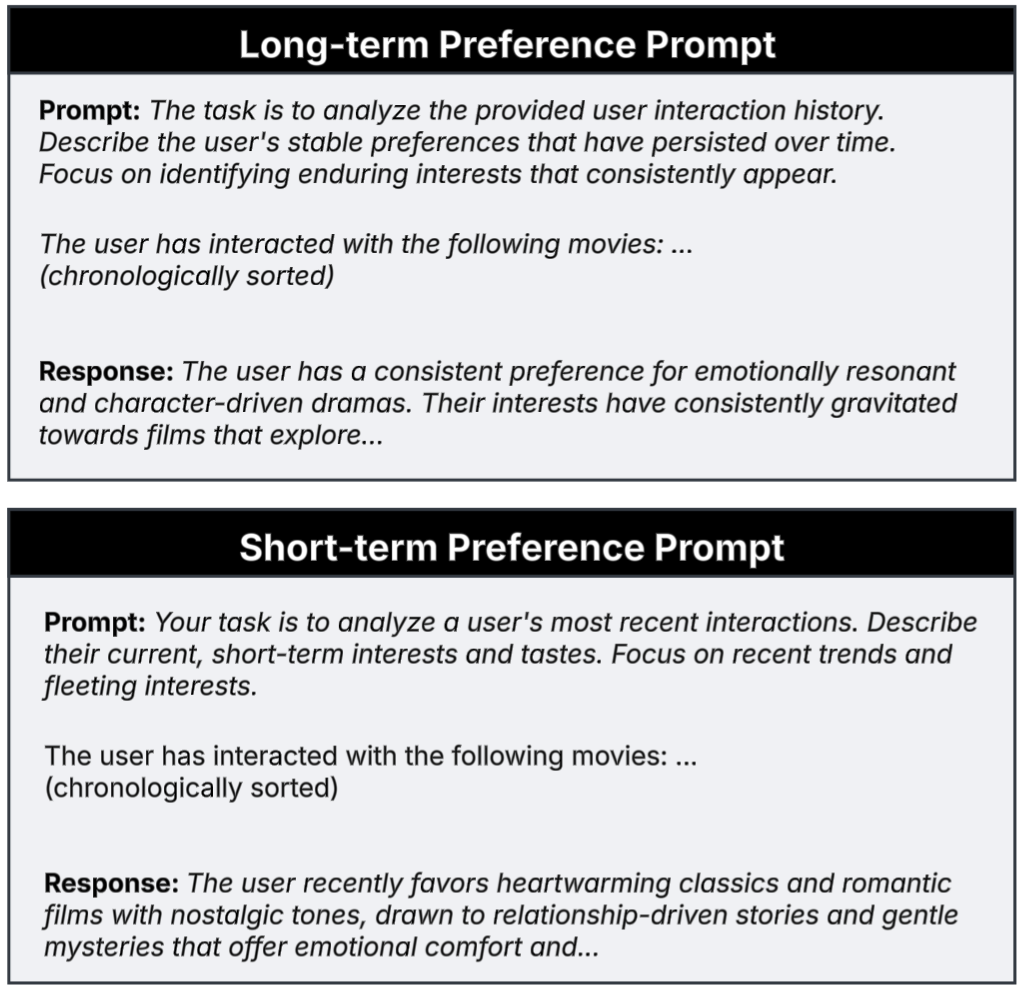}
    \caption{\small  Examples of Prompts for LLM-based Temporal User Profile Generation (Movies\&TV Domain)}
    \label{fig:prompt}
\end{figure}

Let $\mathcal{U} = {u_1, u_2, \ldots, u_{|\mathcal{U}|}}$ be users, and $\mathcal{I} = {i_1, i_2, \ldots\allowbreak, i_{|\mathcal{I}|}}$ be items. Each user $u \in \mathcal{U}$ interacts with a subset of items $\mathcal{I}_u \subseteq \mathcal{I}$, with interactions associated to timestamps. We define the interaction history as $\mathcal{H}u = {(i, t_{u,i}) : i \in \mathcal{I}_u}$ sorted chronologically.

We aim to learn embeddings for users ($\mathbf{e}_u \in \mathbb{R}^d$) and items ($\mathbf{e}_i \in \mathbb{R}^d$) to predict user-item interaction likelihood:
\begin{equation} \label{eq:1} \hat{y}_{u,i} = f(\mathbf{e}_u, \mathbf{e}_i),
\end{equation}
where $f$ is a multi-layer perceptron (MLP).
The pipeline (Figure~\ref{fig:workflow}) comprises three steps: (i) profile generation via LLMs, (ii) semantic embedding, and (iii) interaction prediction.

\textit{(i) LLM-based Temporal Profile Generation}: To capture temporal aspects of user preferences, our approach prompts the LLM twice over each user’s interaction history $\mathcal{H}_u$, using distinct instructions to generate two textual summaries. The first focuses on recent interactions, producing the short-term profile:
\begin{equation} \text{NL}_u^{\text{short}} = \text{LLM}(\mathcal{H}_u, \text{Prompt}^{\text{short}}). \end{equation}
The second summarizes persistent behaviors across the full history, forming the long-term profile:
\begin{equation} \text{NL}_u^{\text{long}} = \text{LLM}(\mathcal{H}_u, \text{Prompt}^{\text{long}}). \end{equation}
These summaries enhance interpretability by separating transient and stable user interests.

To provide a concrete understanding of our LLM-based temporal profile generation, we include a conceptual illustration of the prompts and their corresponding output in Figure \ref{fig:prompt}. The prompts are carefully designed to elicit distinct temporal summaries. As shown, the "Long-term Preference Prompt" emphasizes identifying enduring interests and consistent themes across the user's entire history. In contrast, the "Short-term Preference Prompt" is engineered to focus specifically on recent behaviors and fleeting interests. While the full, detailed prompts and their configurations are available in our public repository (see section \ref{sec:experiments}) to ensure reproducibility, these examples demonstrate the core strategy of using natural language instructions to temporally disentangle user preferences.

\textit{(ii) Semantic Embedding and Attention Fusion}:
Each textual profile is transformed into a high-dimensional embedding via BERT~\cite{devlin-etal-2019-bert}:
\begin{equation}\label{eq:2} \mathbf{r}_u^{\text{short}} = \text{BERT}(\text{NL}_u^{\text{short}}), \end{equation}
\begin{equation}\label{eq:3} \mathbf{r}_u^{\text{long}} = \text{BERT}(\text{NL}_u^{\text{long}}), \end{equation} yielding embeddings that semantically represent short-term and long-term interests, respectively.
To dynamically integrate these temporal embeddings, we apply a learnable attention layer~\cite{waswani2017attention}, producing a unified user embedding:
\begin{equation}\label{eq:4} \alpha_u^{\text{short}} = \frac{\exp(\mathbf{W}_a \mathbf{r}_u^{\text{short}})}{\exp(\mathbf{W}_a \mathbf{r}_u^{\text{short}}) + \exp(\mathbf{W}_a \mathbf{r}_u^{\text{long}})}, \end{equation}
\begin{equation}\label{eq:5} \alpha_u^{\text{long}} = 1 - \alpha_u^{\text{short}}, \end{equation} where $\mathbf{W}_a \in \mathbb{R}^{1 \times d}$ is a learnable parameter vector. The final embedding is:
\begin{equation}\label{eq:6} \mathbf{e}_u = \alpha_u^{\text{short}} \cdot \mathbf{r}_u^{\text{short}} + \alpha_u^{\text{long}} \cdot \mathbf{r}_u^{\text{long}}. \end{equation}

Crucially, the learned weights ($\alpha_u^{\text{short}}$, $\alpha_u^{\text{long}}$) explicitly convey the model’s decision rationale regarding recent versus historical user interests.

\textit{(iii) Interaction Prediction}:
To predict interaction probabilities, we concatenate user embedding $\mathbf{e}_u$ with item embedding $\mathbf{e}_i$ and process through an MLP~\cite{popescu2009multilayer}:
\begin{equation}\label{eq:7} \hat{y}_{u,i} = \text{MLP}\bigl([\mathbf{e}_u; \mathbf{e}_i]\bigr), \end{equation} with output $\hat{y}_{u,i} \in [0,1]$ indicating interaction likelihood. The model is trained via binary cross-entropy loss:

\begin{equation}\label{eq:8}
\mathcal{L}
= -\frac{1}{|\mathcal{D}|}
\sum_{(u,i,y_{u,i}) \in \mathcal{D}}
\Bigl[\, y_{u,i}\log \hat{y}_{u,i}
      + (1-y_{u,i})\log\bigl(1-\hat{y}_{u,i}\bigr) \Bigr].
\end{equation}
where $\mathcal{D}$ is the training set and $y_{u,i}$ the observed interactions.

\begin{table}
\centering
\caption{Datasets Statistics}
\label{tab:dataset_statistics}
\begin{tabular}{lccc}
\toprule
\textbf{Dataset} & \textbf{Users} & \textbf{Items} & \textbf{Interactions} \\
\midrule   
Movies\&TV & 10,000  & 14,420 & 202,583 \\
Games      & 10,371  & 3,790  & 83,842 \\
\bottomrule
\end{tabular}
\end{table}

\begin{table}
\centering
\caption{User Profiles Statistics}
\label{tab:user_statistics}
\begin{tabular}{lcccc}
\toprule
\textbf{Dataset} & \textbf{Mean} & \textbf{Median} & \textbf{Mode} & \textbf{Std Dev} \\
\midrule
Movies\&TV & 11.79 & 9.00 & 6 & 9.80 \\
Games & 4.55  & 3.00 & 3 & 3.97  \\ 
\bottomrule
\end{tabular}
\end{table}

\begin{table*}
    \centering
    \caption{Comparative Performance Evaluation of LLM-Driven Temporal User Profiling Across Different Domains \\ \textit{asterisk denotes a statistically significant improvement over the baseline (Centric) (\(p < 0.05\)).}}
    \label{tab:performance_comparison}
    \resizebox{\textwidth}{!}{ 
    \begin{tabular}{lllllllll}
        \toprule
        \multirow{2}{*}{\textbf{Method}} & \multicolumn{4}{c}{\textbf{Movies\&TV}}  & \multicolumn{4}{c}{\textbf{Video Games}} \\
        \cmidrule(lr){2-5} \cmidrule(lr){6-9} 
        & \textbf{Recall@10} & \textbf{NDCG@10} & \textbf{Recall@20} & \textbf{NDCG@20} & \textbf{Recall@10} & \textbf{NDCG@10} & \textbf{Recall@20} & \textbf{NDCG@20} \\
        \midrule
        Centric & 0.0113 & 0.0191 & 0.0199 & 0.0269 & 0.0645 & 0.0532 & 0.0932 & 0.0649 \\
        Popularity & 0.0082 & 0.0145 & 0.0133 & 0.0191 & 0.0397 & 0.0324 & 0.0706 & 0.0453 \\
        MF & 0.0048 & 0.0085 & 0.0087 & 0.0124 & 0.0457 & 0.0370 & 0.0754 & 0.0491 \\
        Temp-Fusion & \underline{0.0118} & \underline{0.0201} & \underline{0.0207} & \underline{0.0276} & \textbf{0.0693} & \textbf{0.0589} & \underline{0.0982} & \textbf{0.0712} \\
        \textbf{LLM-TP}  & \textbf{0.0132}$^*$ & \textbf{0.0217}$^*$ & \textbf{0.0223}$^*$ & \textbf{0.0293}$^*$ & \underline{0.0665}$^*$ & \underline{0.0547}$^*$ & \textbf{0.1021}$^*$ & \underline{0.0683}$^*$ \\
        \midrule
        Gain of LLM-TP vs. Centric & 17\% & 14\% & 12\% & 9\% & 3\% & 3\% & 10\% & 5\% \\
        \bottomrule
    \end{tabular}
    }
\end{table*}

\section{Experiments} \label{sec:experiments}
We evaluate the effectiveness of the LLM-driven temporal user profiling framework on two Amazon domains~\cite{hou2024bridging}—Movies\&TV and Video Games—chosen for their differing user profile density and behavioral variability. This allows assessment under diverse recommendation conditions. Dataset statistics are in Tables~\ref{tab:dataset_statistics} and~\ref{tab:user_statistics}. For LLM processing, we retain English item descriptions exceeding 500 characters. For benchmarking, we compare against four baselines: \textbf{Centric}~\cite{polignano2021together}, which averages item embeddings without temporal modeling. \textbf{Temp-Fusion} is also included, designed to assess temporal fusion's benefit without LLM-generated semantic profiles; it segments interaction histories into short-term and long-term numerical item embeddings (e.g., by averaging within recent/historical windows) and fuses them via attention, isolating the LLM's textual contribution. Two standard control baselines are \textbf{Popularity}~\cite{ricci2011recommender}, a non-personalized ranking approach, and \textbf{Matrix Factorization (MF)}~\cite{koren2009matrix}, a collaborative filtering technique without textual features. Evaluation is performed using a rigorous per-user temporal holdout protocol. For each user, we first chronologically sort their interactions by timestamp. The dataset is then split so that a user's training data always precedes their validation and test data. This ensures that our model is trained only on past behaviors to predict future ones, thereby strictly preventing within-user temporal data leakage. We report model performance using standard metrics, including Recall@K and NDCG@K. Textual summaries are encoded via SBERT~\cite{reimers-2019-sentence-bert} (MiniLM-L6-v2, 384-dim), with profiles generated using GPT-40-mini~\cite{achiam2023gpt}. Interaction likelihoods are predicted using an MLP (hidden size 128, dropout 0.2) trained with binary cross-entropy, batch size 2048, Adam optimizer, and early stopping (patience=5), executed on four NVIDIA A100 GPUs. All code, data, and prompts are available in a public GitHub repository\footnote{https://github.com/milsab/UMRec}.

\subsection{Results and Discussion}\label{subsubsec:discussion}

This section presents the empirical results of our assessment, summarized in Table \ref{tab:performance_comparison}, highlighting key insights into the capability of the LLM-driven temporal user profiling approach.

\textbf{Temporal Profiling Benefits Domains with High User Activity:} On the Movies\&TV dataset, characterized by larger and higher user activity (average profile size: 11.79, std: 9.80), the evaluated framework demonstrates significant effectiveness. It notably outperforms the Centric baseline (17\% Recall@10, 14\% NDCG@10 improvements) and Temp-Fusion. These observations underscore the value of incorporating semantically rich, natural-language user profiles from LLMs, particularly where user behaviors are varied and frequently changing.

\textbf{Mixed Effectiveness for Domains with Lower User Activity:} In the Video Games domain, with smaller and less frequent user interactions (mean: 4.55, std: 3.97), the assessment reveals mixed effectiveness. While our approach achieves the highest Recall@20, Temp-Fusion slightly surpasses it at smaller K-values. This outcome likely arises from interaction sparsity; our conjecture is that lower user activity often leads to comparatively more stable preferences, limiting the incremental advantage of detailed temporal modeling through textual differentiation. This highlights a critical trade-off between enhanced predictive power and associated computational costs, suggesting context-dependent LLM application.

\textbf{Intrinsic Explainability Potential:} The framework inherently supports interpretability (Figure~\ref{fig:case_study}). LLM-generated natural language summaries (Figure~\ref{fig:case_study}A and ~\ref{fig:case_study}B) clearly reflect distinct short-term versus long-term user interests, providing human-readable profiles. Additionally, learned attention weights ($\alpha_u^{\text{short}}$, $\alpha_u^{\text{long}}$) transparently indicate the relative influence of these profiles in the unified user embedding, aiding understanding of recommendation rationale. Figure~\ref{fig:case_study}C illustrates how these components could generate user-facing explanations, an aspect for future work to fully realize built-in transparency.

In summary, our experiments suggest that temporal profiling with LLMs is practical and interpretable, especially for domains with dynamic user behavior. Future work will focus on developing user-facing interfaces and evaluating explanations to enhance transparency and trust.

\begin{figure}
    \centering
    \includegraphics[height=0.25\textheight]{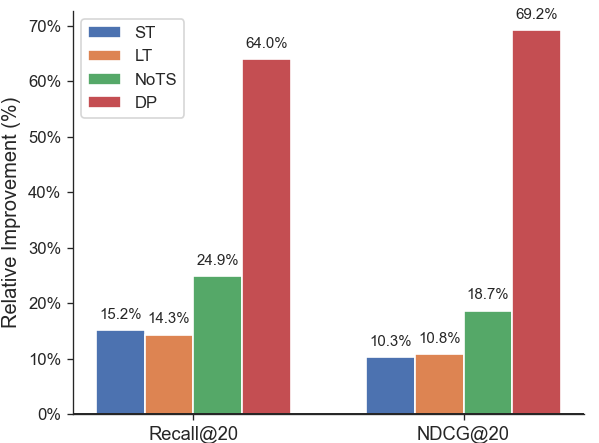}
    \caption{\small Relative gains in Recall@20 and NDCG@20 of the full model over ablation variants on the Movies\&TV dataset.}
    \label{fig:bar_chart}
\end{figure}

\section{Ablation Study} \label{sec:ablation}

We conducted an ablation study to determine the impact of individual components within the evaluated framework on recommendation accuracy, using the Movies\&TV dataset. Table~\ref{tab:ablation_study} summarizes the performance metrics, and Fig.~\ref{fig:bar_chart} highlighting relative improvements. More results are in the GitHub repository (see Section \ref{sec:experiments}). The ablation results underscore the importance of temporal modeling and non-linear scoring. The \textbf{General Preferences (No TS)} variant, where the LLM generated a single, holistic user profile (distinct from Temp-Fusion's numerical aggregation), showed a consistent performance drop of over 20\% in Recall and 18\% in NDCG, emphasizing the value of LLM-generated temporal distinction (Fig.~\ref{fig:bar_chart}). The full framework also consistently outperformed Short-Term Only (ST) and Long-Term Only (LT) variants individually (e.g., +15.2\% Recall@20 over ST), confirming complementary signals from both temporal components. Lastly, replacing MLP scoring with a dot product (DP variant) resulted in the steepest degradation, highlighting the necessity of non-linear interaction modeling. Overall, Figure~\ref{fig:bar_chart} and Table~\ref{tab:ablation_study} demonstrate that temporal disentanglement via LLM-based profile summarization and non-linear embedding fusion collectively contribute to the observed robust recommendation performance within the evaluated framework.

\begin{table}[t]
\centering
\caption{Comparison for Ablation Variants (Movies\&TV)}
\label{tab:ablation_study}
\begin{tabular}{lcc}
\hline
\textbf{Ablation Variant} & \textbf{Recall@20} & \textbf{NDCG@20} \\
\hline
Short-Term Only (ST) & 0.0193 & 0.0266 \\
Long-Term Only (LT) & 0.0195 & 0.0265 \\
General Preferences (No TS) & 0.0178 & 0.0247 \\
Dot-Product Scoring (DP) & 0.0136 & 0.0173 \\
\textbf{Full Model (LLM-TP)} & \textbf{0.0223} & \textbf{0.0293} \\
\hline
\end{tabular}
\end{table}

\section{Conclusion}
This paper presented an exploratory evaluation of a lightweight framework that leverages LLM-driven temporal user profiling within a pipeline demonstrating interpretability potential. Our evaluation indicates varied capabilities: notable gains in high-activity domains (e.g., Movies\&TV) versus less pronounced benefits in sparser environments (e.g., Video Games). This highlights a critical trade-off between enhanced performance and computational costs, suggesting context-dependent LLM application. The framework inherently offers a promising pathway toward transparency through its natural language profiles and attention weights, providing an intrinsic potential for interpretability. These early insights contribute to understanding the practical capability of LLM-driven temporal user profiling. Future work includes developing user-facing explanations, conducting user studies, and exploring adaptive LLM strategies.


\begin{thebibliography}{10}
\providecommand{\url}[1]{#1}
\csname url@samestyle\endcsname
\providecommand{\newblock}{\relax}
\providecommand{\bibinfo}[2]{#2}
\providecommand{\BIBentrySTDinterwordspacing}{\spaceskip=0pt\relax}
\providecommand{\BIBentryALTinterwordstretchfactor}{4}
\providecommand{\BIBentryALTinterwordspacing}{\spaceskip=\fontdimen2\font plus
\BIBentryALTinterwordstretchfactor\fontdimen3\font minus \fontdimen4\font\relax}
\providecommand{\BIBforeignlanguage}[2]{{%
\expandafter\ifx\csname l@#1\endcsname\relax
\typeout{** WARNING: IEEEtran.bst: No hyphenation pattern has been}%
\typeout{** loaded for the language `#1'. Using the pattern for}%
\typeout{** the default language instead.}%
\else
\language=\csname l@#1\endcsname
\fi
#2}}
\providecommand{\BIBdecl}{\relax}
\BIBdecl

\bibitem{devlin-etal-2019-bert}
\BIBentryALTinterwordspacing
J.~Devlin, M.-W. Chang, K.~Lee, and K.~Toutanova, ``{BERT}: Pre-training of deep bidirectional transformers for language understanding,'' in \emph{Proceedings of the 2019 Conference of the North {A}merican Chapter of the Association for Computational Linguistics: Human Language Technologies, Volume 1 (Long and Short Papers)}, J.~Burstein, C.~Doran, and T.~Solorio, Eds.\hskip 1em plus 0.5em minus 0.4em\relax Minneapolis, Minnesota: Association for Computational Linguistics, Jun. 2019, pp. 4171--4186. [Online]. Available: \url{https://aclanthology.org/N19-1423/}
\BIBentrySTDinterwordspacing

\bibitem{waswani2017attention}
A.~Waswani, N.~Shazeer, N.~Parmar, J.~Uszkoreit, L.~Jones, A.~Gomez, L.~Kaiser, and I.~Polosukhin, ``Attention is all you need,'' in \emph{NIPS}, 2017.

\bibitem{polignano2021together}
M.~Polignano, C.~Musto, M.~de~Gemmis, P.~Lops, and G.~Semeraro, ``Together is better: Hybrid recommendations combining graph embeddings and contextualized word representations,'' in \emph{Proceedings of the 15th ACM conference on recommender systems}, 2021, pp. 187--198.

\bibitem{koren2009collaborative}
Y.~Koren, ``Collaborative filtering with temporal dynamics,'' in \emph{Proceedings of the 15th ACM SIGKDD international conference on Knowledge discovery and data mining}, 2009, pp. 447--456.

\bibitem{hidasi2015session}
B.~Hidasi, A.~Karatzoglou, L.~Baltrunas, and D.~Tikk, ``Session-based recommendations with recurrent neural networks,'' \emph{arXiv preprint arXiv:1511.06939}, 2015.

\bibitem{kang2018self}
W.-C. Kang and J.~McAuley, ``Self-attentive sequential recommendation,'' in \emph{2018 IEEE international conference on data mining (ICDM)}.\hskip 1em plus 0.5em minus 0.4em\relax IEEE, 2018, pp. 197--206.

\bibitem{chi2022long}
H.~Chi, H.~Xu, H.~Fu, M.~Liu, M.~Zhang, Y.~Yang, Q.~Hao, and W.~Wu, ``Long short-term preference modeling for continuous-time sequential recommendation,'' \emph{arXiv preprint arXiv:2208.00593}, 2022.

\bibitem{zhang2020explainable}
Y.~Zhang, X.~Chen \emph{et~al.}, ``Explainable recommendation: A survey and new perspectives,'' \emph{Foundations and Trends{\textregistered} in Information Retrieval}, vol.~14, no.~1, pp. 1--101, 2020.

\bibitem{zhao2023survey}
W.~X. Zhao, K.~Zhou, J.~Li, T.~Tang, X.~Wang, Y.~Hou, Y.~Min, B.~Zhang, J.~Zhang, Z.~Dong \emph{et~al.}, ``A survey of large language models,'' \emph{arXiv preprint arXiv:2303.18223}, 2023.

\bibitem{chang2024survey}
Y.~Chang, X.~Wang, J.~Wang, Y.~Wu, L.~Yang, K.~Zhu, H.~Chen, X.~Yi, C.~Wang, Y.~Wang \emph{et~al.}, ``A survey on evaluation of large language models,'' \emph{ACM transactions on intelligent systems and technology}, vol.~15, no.~3, pp. 1--45, 2024.

\bibitem{zhao2024recommender}
Z.~Zhao, W.~Fan, J.~Li, Y.~Liu, X.~Mei, Y.~Wang, Z.~Wen, F.~Wang, X.~Zhao, J.~Tang \emph{et~al.}, ``Recommender systems in the era of large language models (llms),'' \emph{IEEE Transactions on Knowledge and Data Engineering}, 2024.

\bibitem{wu2024survey}
L.~Wu, Z.~Zheng, Z.~Qiu, H.~Wang, H.~Gu, T.~Shen, C.~Qin, C.~Zhu, H.~Zhu, Q.~Liu \emph{et~al.}, ``A survey on large language models for recommendation,'' \emph{World Wide Web}, vol.~27, no.~5, p.~60, 2024.

\bibitem{lubos2024llm}
S.~Lubos, T.~N.~T. Tran, A.~Felfernig, S.~Polat~Erdeniz, and V.-M. Le, ``Llm-generated explanations for recommender systems,'' in \emph{Adjunct Proceedings of the 32nd ACM Conference on User Modeling, Adaptation and Personalization}, 2024, pp. 276--285.

\bibitem{ma2024xrec}
Q.~Ma, X.~Ren, and C.~Huang, ``Xrec: Large language models for explainable recommendation,'' \emph{arXiv preprint arXiv:2406.02377}, 2024.

\bibitem{guo2020survey}
Q.~Guo, F.~Zhuang, C.~Qin, H.~Zhu, X.~Xie, H.~Xiong, and Q.~He, ``A survey on knowledge graph-based recommender systems,'' \emph{IEEE Transactions on Knowledge and Data Engineering}, vol.~34, no.~8, pp. 3549--3568, 2020.

\bibitem{wang2025knowledge}
S.~Wang, W.~Fan, Y.~Feng, S.~Lin, X.~Ma, S.~Wang, and D.~Yin, ``Knowledge graph retrieval-augmented generation for llm-based recommendation,'' \emph{arXiv preprint arXiv:2501.02226}, 2025.

\bibitem{shimizu2022explainable}
R.~Shimizu, M.~Matsutani, and M.~Goto, ``An explainable recommendation framework based on an improved knowledge graph attention network with massive volumes of side information,'' \emph{Knowledge-Based Systems}, vol. 239, p. 107970, 2022.

\bibitem{shi2024llm}
G.~Shi, X.~Deng, L.~Luo, L.~Xia, L.~Bao, B.~Ye, F.~Du, S.~Pan, and Y.~Li, ``Llm-powered explanations: Unraveling recommendations through subgraph reasoning,'' \emph{arXiv preprint arXiv:2406.15859}, 2024.

\bibitem{tan2016improved}
Y.~K. Tan, X.~Xu, and Y.~Liu, ``Improved recurrent neural networks for session-based recommendations,'' in \emph{Proceedings of the 1st workshop on deep learning for recommender systems}, 2016, pp. 17--22.

\bibitem{zhu2017next}
Y.~Zhu, H.~Li, Y.~Liao, B.~Wang, Z.~Guan, H.~Liu, and D.~Cai, ``What to do next: Modeling user behaviors by time-lstm.'' in \emph{IJCAI}, vol.~17, 2017, pp. 3602--3608.

\bibitem{popescu2009multilayer}
M.-C. Popescu, V.~E. Balas, L.~Perescu-Popescu, and N.~Mastorakis, ``Multilayer perceptron and neural networks,'' \emph{WSEAS Transactions on Circuits and Systems}, vol.~8, no.~7, pp. 579--588, 2009.

\bibitem{hou2024bridging}
Y.~Hou, J.~Li, Z.~He, A.~Yan, X.~Chen, and J.~McAuley, ``Bridging language and items for retrieval and recommendation,'' \emph{arXiv preprint arXiv:2403.03952}, 2024.

\bibitem{ricci2011recommender}
F.~Ricci, L.~Rokach, and B.~Shapira, ``Introduction to recommender systems handbook,'' in \emph{Recommender systems handbook}.\hskip 1em plus 0.5em minus 0.4em\relax Springer, 2011, pp. 1--35.

\bibitem{koren2009matrix}
Y.~Koren, R.~Bell, and C.~Volinsky, ``Matrix factorization techniques for recommender systems,'' \emph{Computer}, vol.~42, no.~8, pp. 30--37, 2009.

\bibitem{reimers-2019-sentence-bert}
\BIBentryALTinterwordspacing
N.~Reimers and I.~Gurevych, ``Sentence-bert: Sentence embeddings using siamese bert-networks,'' in \emph{Proceedings of the 2019 Conference on Empirical Methods in Natural Language Processing}.\hskip 1em plus 0.5em minus 0.4em\relax Association for Computational Linguistics, 11 2019. [Online]. Available: \url{https://arxiv.org/abs/1908.10084}
\BIBentrySTDinterwordspacing

\bibitem{achiam2023gpt}
J.~Achiam, S.~Adler, S.~Agarwal, L.~Ahmad, I.~Akkaya, F.~L. Aleman, D.~Almeida, J.~Altenschmidt, S.~Altman, S.~Anadkat \emph{et~al.}, ``Gpt-4 technical report,'' \emph{arXiv preprint arXiv:2303.08774}, 2023.

\end{thebibliography}


\end{document}